# A 140 Mpc Oscillation of the Abell Cluster Correlation Function

Veikko Saar[1,2]
*Tartu University, EE–2400 Tartu, Estonia*

Erik Tago, Jaan Einasto[2], Maret Einasto
*Tartu Astrophysical Observatory, EE–2444 Tõravere, Estonia*

and

Heinz Andernach[3]
*Observatoire de Lyon, 9 Ave. Charles Andre, F–69561, St-Genis-Laval Cedex, France*

## ABSTRACT

We have developed a parameter–independent method to detect local maxima of the two–point correlation function. By applying it to two samples of rich Abell clusters of galaxies with redshift limits $z < 0.08$ and $z < 0.12$ we detect three maxima centered at $150\ h^{-1}$ Mpc, $300\ h^{-1}$ Mpc and $430\ h^{-1}$ Mpc with confidence levels 80% and higher. This sequence of fluctuations has an average interval of $140\ h^{-1}$ Mpc, that can be explained by a power spectrum with a distinct peak at $k = 0.048 \pm 0.005\ h$ Mpc$^{-1}$.

*Subject headings:* cosmology; observations – galaxies: clustering – large-scale structure of the universe – methods: numerical

---

[1]Tartu Astrophysical Observatory, EE–2444 Tõravere, Estonia
[2]Max–Planck–Institut für Extraterrestrische Physik, 85740 Garching, Germany
[3]Present address: IUE Observatory, Villafranca, Apartado 50727, E-28080 Madrid, Spain

## 1. Introduction

There has been a growing evidence on the presence of large-scale inhomogeneities in the distribution of matter at scales around 100 $h^{-1}$ Mpc. The first hints came from the distribution of very rich clusters of galaxies, observed by Kopylov *et al.* (1984, 1988) and later by Fetisova *et al.* (1993). They found a secondary maximum in the two-point correlation function that has been later confirmed by Mo *et al.* (1992a, b) and Einasto & Gramann (1993). However, the statistical significance of the secondary maximum in the cluster correlation function was never assessed.

Deep pencil-beam surveys have revealed density peaks in the distribution of galaxies with a surprisingly regular spacing of about 128 $h^{-1}$ Mpc (Broadhurst *et al.* 1990). These peaks were attributed to the locations of superclusters as defined by rich clusters of galaxies (Bahcall 1991). The power spectrum of the density distribution in this one-dimensional survey has a sharp maximum at about 130 $h^{-1}$ Mpc. Whether this feature is present in three-dimensional space is less clear, since redshift surveys of galaxies up to $z \approx 0.2$, covering a large fraction of the sky, are only in the planning stage (Gunn & Weinberg 1995). However, an analysis of the three-dimensional distribution of superclusters suggests indeed the presence of the characteristic scale at about 130 $h^{-1}$ Mpc in the supercluster-void network (Einasto *et al.* 1994).

Such a scale, if present, should also be reflected in the power spectrum. The power spectrum of matter has a positive index $n \approx 1$ on very large scales as suggested by theoretical arguments and confirmed by COBE (Smoot *et al.* 1992), and a negative index $-2 < n < -1$ on small scales, as emphasized by recent studies of the distribution of galaxies of various types. Thus there must be a transition on intermediate scales which can be identified as a peak. The CfA Redshift Survey of galaxies shows a turnover at $\lambda \approx 150$ $h^{-1}$ Mpc, albeit with a small statistical significance (Vogeley *et al.* 1992). A comparison of IRAS galaxies with Abell clusters and radio galaxies suggests a maximum at 160 $h^{-1}$ Mpc (Mo, Peacock & Xia 1992). The maximum is also detected in the distribution of rich Abell clusters ($R \geq 1$) at wavelengths $\lambda > 100$ $h^{-1}$ Mpc (Peacock & West 1992, hereafter PW), as well as in the whole Abell/ACO cluster sample ($R \geq 0$) (Einasto *et al.* 1993). In contrast, no turnover has been found in the recent CfA Redshift Survey, though the spectrum flattens at $\lambda > 120$ $h^{-1}$ Mpc (Park *et al.* 1994). As the shape of the spectrum has an influence on the matter-void network (Frisch *et al.* 1995), it should be possible to use the distribution of objects on large scales to determine the behavior of the power spectrum for these wavelengths.

Although most of these results support the presence of large-scale inhomogeneities, the statistical significance of this feature is unclear. The order of the characteristic scale found so far is in agreement with pencil-beam surveys and can be expected at about 130 $h^{-1}$ Mpc (Dekel *et al.* 1992).

The purpose of this paper is to investigate the presence of inhomogeneities in three-dimensional density distributions and to estimate the statistical significance of the result. We shall base our analysis on the Abell/ACO catalog of rich clusters of galaxies which is the deepest available complete survey covering a large fraction of the sky. The paper is organized as follows: in section 2 we describe the observational data used, in section 3 we develop a novel method for the detection of maxima of the two-point correlation function, in section 4 we present our analysis and the main results are summarized in the conclusions.

## 2. Data

We shall base our analysis on the Abell/ACO catalog of rich clusters of galaxies (Abell 1958; Abell, Corwin & Olowin 1989). The redshift information for these clusters has been continuously updated by two of us (ET & HA; cf. Andernach *et al.* 1994).

It has been pointed out that this catalog is biased by projection effects (Sutherland & Efstathiou 1991). However, this effect is present mostly in the sample that includes all clusters ($R \geq 0$), and absent for the $R \geq 1$ subsample, for which there is no evidence that the close pairs of clusters reflect anything other than true spatial correlations (PW).

We shall first consider the same sample used by Peacock & West (1992), but exclude low galactic latitudes to ensure better completeness for higher redshifts. The limiting criteria are:

$$R \geq 1, \quad 0.01 < z < 0.08,$$
$$b > 40°, \quad b < -30°.$$

This selection includes 175 clusters with measured redshifts and 3 clusters with estimated redshifts. For the estimation of photometric redshifts we used the formulae proposed by Peacock & West (1992).

We shall also consider a sample with the same selection parameters but using a higher redshift limit. Since the catalog becomes largely incomplete in measured redshifts at about $z \approx 0.13$, we limit the second sample to $0.01 < z < 0.12$. This includes 382 measured redshifts and 111 estimated ones, giving about 78% for completeness in redshifts.

The number density of clusters within $0.01 < z < 0.08$ is 1.6 times higher than within $0.08 < z < 0.12$, indicating that there might be present a selection effect.

### 2.1. Uncertainties

The uncertainties in deriving the cluster distances from their redshifts can be roughly divided into two categories – those for clusters with measured redshifts (the main bulk of the data) and those for clusters with distances estimated from their luminosity function.

For clusters with measured redshifts the highest uncertainties are caused by peculiar velocities, both of galaxies within a cluster and of clusters as a whole. The uncertainties caused by peculiar velocities of galaxies can be corrected to some extent by using redshift measurements of different galaxies in the same cluster. This will reduce the errors by a factor of $\sqrt{N_z}$, where $N_z$ is the number of measured redshifts in the particular cluster. The average velocity dispersion has been found between 657 km/s for $R \geq 0$, 758 km/s for $R \geq 1$ and 989 km/s for $R \geq 2$ clusters (Girardi et al. 1993).

Another error of the same order is due to the peculiar velocities of clusters. The Local Group has been found to move at 600 km/s relative to the Hubble flow (Rowan–Robinson et al. 1990). An analysis of nearby Abell clusters gave an average velocity dispersion of $600 \pm 105$ km/s (Postman, Huchra & Geller 1992). On the basis of these results, we use an error estimate

$$\sigma_c = 600 \text{ km/s}, \; \sigma_g = 800 \text{ km/s},$$
$$\sigma_v^2 = \sigma_c^2 + \sigma_g^2/N_z. \quad (1)$$

Additional uncertainties arise from the spurious assignment of redshifts of foreground or background galaxies to clusters. One can assume these clusters to have large discrepancies $|z_{est} - z_{obs}|$ and small $N_z$. We found 21 clusters affecting our samples which had measured redshifts more than twice different from the estimated ones, 19 of which had measured redshifts less than the estimates. For these clusters we use only the estimated redshifts. The numbers of clusters given above already include this correction.

The error in estimated redshifts is much larger, and is found to be about 27% for the northern sample and 18% for the south (PW). Since the magnitude-redshift relation is determined in the $ln\ z \sim m$ scale, the errors should have a normal distribution relative to $ln\ z$. We built the regression of $ln\ z_{obs}$ on $ln\ z_{est}$ for $R \geq 1$ clusters in the range $0.01 < z < 0.12$. The rms error of the regression was

$$\sigma(ln\ z) = 0.27. \quad (2)$$

This leads to slightly larger deviations towards longer distances (31%) than towards us (24%).

### 3. Method

The two–point correlation function is traditionally determined by binning the distribution of distances between sample objects. This leads to a dependence of the results on the chosen bin size. Even if the average correlation function $\xi$ can point out general trends, the standard error estimate depends on the width of the bin and should not be used (see Fig. 1).

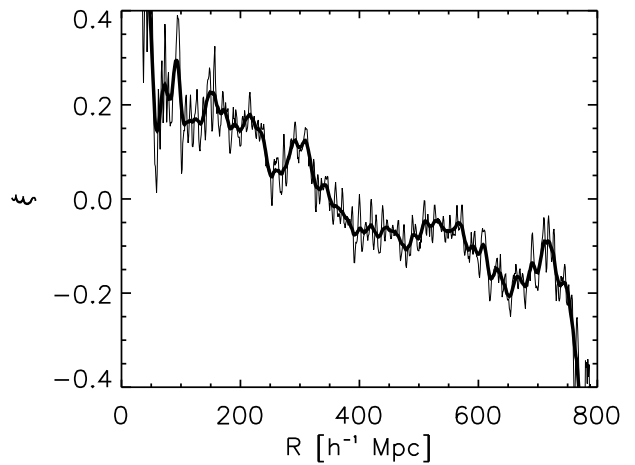

Fig. 1.— Two–point correlation function of the larger sample ($0.01 < z < 0.12$). The thin line shows the convolution with a 5 $h^{-1}$ Mpc triangular kernel, the thick line is the convolution with a 20 $h^{-1}$ Mpc kernel. The negative trend of $\xi$ is probably due to the incompleteness of the cluster catalog at high redshifts (see Fig. 2). The dependence of deviations on the smoothing scale is evident.

However, there exists a precise function of the data that does not depend on arbitrary parameters,

namely the cumulative distribution of distances of cluster pairs. It has the disadvantage of being not illustrative and it is not clear how to find inhomogeneities in this distribution.

We estimate the correlation function as

$$\xi(r) = dP_{cc}(r)/dP_{rr}(r) - 1, \quad (3)$$

where $dP_{cc}$ is the probability density of distances of cluster pairs and $dP_{rr}$ is the probability density of a Poisson distribution in the same volume. This gives us an expression for the cumulative distribution of distances of cluster pairs:

$$P_{cc}(L) = \int_0^L (\xi(r) + 1) dP_{rr}(r). \quad (4)$$

Since our data consist of discrete points, the integral changes to a sum that can be done in an arbitrary interval:

$$\sum_{r=L1}^{r=L2} \Delta P_{cc}(r) = \sum_{r=L1}^{r=L2} (\xi(r) + 1) \Delta P_{rr}(r). \quad (5)$$

The left–hand side of this equation is independent of binning as it just counts all pairs with mutual distances between $L1$ and $L2$. The term $\Delta P_{rr}$ does the same on the right–hand side, but it has an additional factor $\xi(r)$. Thus in order to use the cumulative distribution function we have to parametrize the two–point correlation function. This might be difficult to do for the full range of distances, but is certainly applicable for a selected range. As we wish to detect maxima of the correlation function, we use equation (5) for a limited distance interval and approximate the behavior of $\xi$ in the interval with a parabola. This leaves no binning–dependent terms in the formula.

The procedure goes as follows. We fix the range $(L1, L2)$. If $\xi$ has a maximum somewhere between $L1$ and $L2$, then the least–squares fit gives a parabola with a maximum in this range. There could also be a minimum within the range. If there is neither maximum nor minimum, the parabola has an extremum outside of the selected range. We shall also check the goodness of the fit, since the parabola might be a bad approximation (eg., if there are more than one maxima). The criterion for this is $\sigma_{cr} = 1/N_p$, where $N_p$ is the number of distances in the chosen interval. As $\sigma_{cr}$ is the increase of $P_{cc}$ per cluster pair, fits giving a residual dispersion $\sigma$ greater than $\sigma_{cr}$ have higher deviations than the signal and will be discarded.

We check all possible combinations of $L1$ and $L2$, looking for local maxima at all intervals and at all scales within the sample volume. For each interval we perform a bootstrap resampling (Efron 1979; Efron & Tibshirani 1986) of our cluster catalog by selecting clusters randomly with replacement. This procedure will usually select about 1/3 of the clusters more than once, and while it is acceptable for many statistics, the repeated clusters cannot be used to estimate the distance distribution of cluster pairs. One possible solution is to smooth the observed distance distribution and to resample distances from the smoothed distribution instead of resampling clusters (Silverman & Young 1987). Another one is to continue resampling from the original data set, perturbing positions of the sampled clusters (Kendall & Kendall 1980). We did not use either of these methods. The first possibility may give a too small variance and the chosen set of pairs would not correspond to any particular distribution of clusters. For the second one we did not find a smoothing distribution that would not induce a smearing scale in the correlation function. We decided to count repeated clusters as being single, that gave us sample sizes about 2/3 of the original dataset. The average distance distribution of these samples coincides with the observed distance distribution, and the result of our choice could only be overestimating the variance of the distribution that is at least not misleading.

After selecting the subsample, we shift clusters by distance uncertainties selected from the normal distribution (see eq. [1]) for clusters with measured redshifts and from the lognormal distribution with the rms error given by eq. [2] for clusters with estimated redshifts. For each bootstrap sample we perform the analysis described above. The bootstrap resampling is repeated 1000 times for each interval and fits with an average $\sigma$ greater than $\sigma_{cr}$ are discarded.

## 4. Analysis

Our method is robust in the sense that it does not depend on the magnitude of $\xi$, but only on its local oscillations. This gives us the freedom not to correct for any global trends, such as selection effects. In order to see how a possible decrease of the cluster number density with distance changes the correlation function, we calculated $\xi$ for a sphere in which the density changed linearly with radius, falling to half its central value at the edges of the sphere (Fig. 2).

The result shows clearly that this causes a large-scale trend without local fluctuations and may induce a maximum within an interval that is about a half of the size of the whole volume.

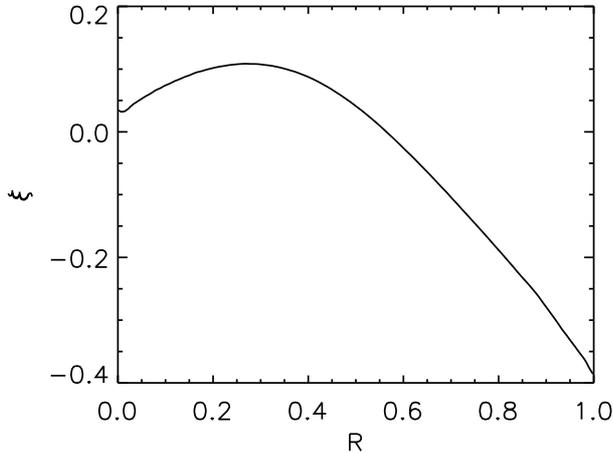

Fig. 2.— Correlation function of a simulated selection effect within a sphere with diameter $D = 1$. The matter density falls to half its central value at the edges of the sphere.

Let us first consider possible results. We may find no maximum in the selected interval, in which case we can not say if it is missing or if our method is simply too insensitive to detect it. Secondly, there may be a maximum detected at a very low confidence level (around 50%), in which case it is caused by few clusters, thus being a feature of the particular sample. Thirdly, if a maximum is found with a confidence level near 100%, then it is not sensitive to local variations and should exist in all samples, thus in both of the cluster samples.

We shall first consider the nearby sample that extends to 520 $h^{-1}$ Mpc. As the zero-point of the correlation function has been found around 50 $h^{-1}$ Mpc (Klypin & Rhee 1994), we expect no secondary maxima at smaller distances. We change the limits of the interval with a step of 5 $h^{-1}$ Mpc and plot the positions of maxima for all intervals with confidence levels 50% and higher against the search interval ($L2 - L1$) in Fig. 3. This plot shows a lot of substructure but it is evident that there is at least one maximum between 100 and 200 $h^{-1}$ Mpc. The highest confidence level is 98% for a maximum centered at 135 $h^{-1}$ Mpc on a 160 $h^{-1}$ Mpc search interval. The second highest confidence level is 83% for a maximum at 440 $h^{-1}$ Mpc on an interval of 120 $h^{-1}$ Mpc.

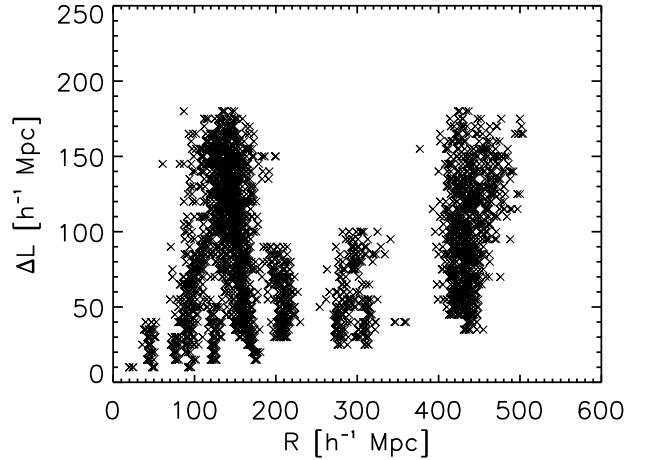

Fig. 3.— Positions of maxima for search intervals with confidence level higher than 50% for the smaller sample, plotted against the width of the interval. This scatter diagram shows the distribution of maxima within all selected intervals.

The volume of the second sample is about 4 times larger than that of the first, so we expect it to hide the local features of the smaller sample. Unfortunately about 1/5-th of the clusters have redshift uncertainties around 30% that may make some maxima undetectable. The results are given in Fig. 4. The confidence level of the most significant maxima in the first sample has lowered and is 89%, perhaps due to the inclusion of clusters with high distance uncertainties. The highest confidence level in this sample is 94% for a maximum at 300 $h^{-1}$ Mpc on intervals of $100 - 120$ $h^{-1}$ Mpc. The average position of the first maximum has shifted to 150 $h^{-1}$ Mpc . The confidence level of the third maximum is 72% at 425 $h^{-1}$ Mpc on a 100 $h^{-1}$ Mpc search interval. There seems to be a fourth maximum centered at 515 $h^{-1}$ Mpc, but since it has a confidence level of only 56%, we shall not consider it in the analysis. An important feature of the maxima is that they extend from intervals close to zero up to certain limits $\Delta L_{max}$, thus being independent on the width of the search interval.

There is no doubt that the two strongest maxima are not local features of the samples. The presence of smaller maxima is less clear, but those that show up in both samples may well be present in general. There is no sign of maxima at larger scales — the widest maxima we found had a width of about 220 $h^{-1}$ Mpc, and the check was done up to scales of 800 $h^{-1}$ Mpc. The results also show a periodic sequence of maxima with a step of about 140 $h^{-1}$ Mpc and a possible sequence of weaker maxima with about twice shorter wavelength. As we can not check whether the weaker maxima are also repeated at the positions of the main sequence or whether they occur only between the main maxima, we shall base our conclusions mainly on the sequence of maxima with highest amplitudes.

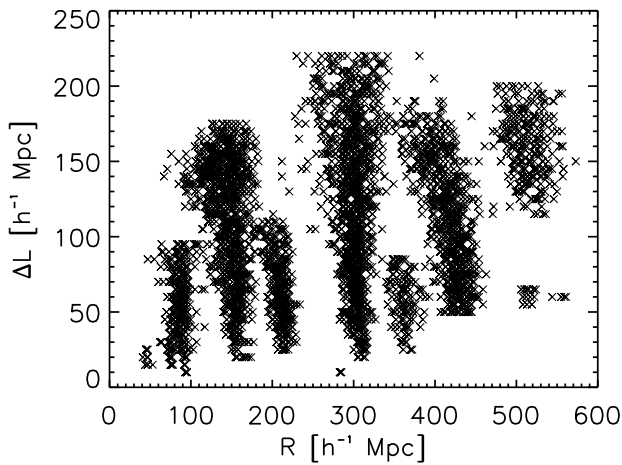

Fig. 4.— The scatter diagram shows positions of maxima for search intervals with confidence level higher than 50% for the larger sample. No maxima were found outside of the ranges of the plot.

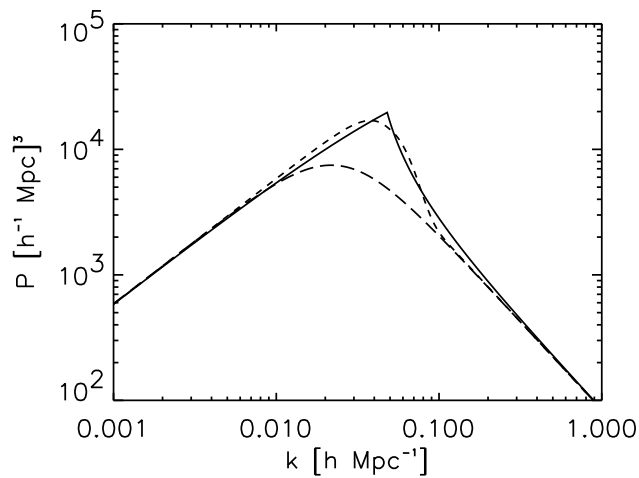

Fig. 5.— Power spectra fixed on small and large scales. The long-dashed line is the fit of Peacock & West (1992), the short-dashed line is for a smooth and the solid line for a sharp maximum at $0.048\ h$ Mpc$^{-1}$.

It is interesting to note that a similar behavior of the two-point correlation function can be predicted by a simple model of the power spectrum. We take the form of the power spectrum used by Peacock & West (1992),

$$P(k) = Ak/(1 + (k/k_0)^{1-n}), \qquad (6)$$

where the fit to Abell clusters gave $k_0 = 0.025\ h$ Mpc$^{-1}$ and $n = -1.4$ (PW). The correlation function can be found by

$$\xi(r) = 4\pi \int_0^\infty P(k)\ k^2 \frac{\sin kr}{kr}\ dk \qquad (7)$$

and is plotted in Fig. 6. It does not exhibit the repetitive pattern, but fluctuations can be produced by using the same spectrum with a sharper maximum. By fixing the small- and large-scale end of the spectrum to match the fit (PW), the period of the oscillation is given by the position of the maximum as $\lambda = 2\pi/k_m$ if the peak is sharp. As shown on Fig. 5-6, the most suitable spectrum has a distint peak. Smoother maxima give either smaller amplitudes or no oscillations at all, not depending on the height. The position of the peak is restricted to $0.043 < k_m < 0.053\ h$ Mpc$^{-1}$ by the data.

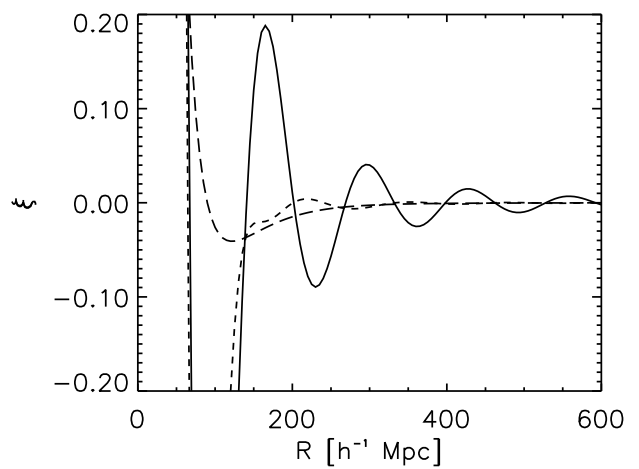

Fig. 6.— Correlation functions corresponding to power spectra on Fig. 5. The notation is the same as used on the previous figure. Notice how strong effect the sharpness of the maximum of the power spectrum has on the fluctuations of $\xi$.

Since the amplitude of the sequence of fluctuations decreases with distance, it may be impossible to detect maxima at very large separations. This agrees with our results as we did not find any maxima at $R > 600\ h^{-1}$ Mpc.

## 5. Conclusions

We have found a sequence of maxima in the two-point correlation function of rich clusters of galaxies which can not be explained by peculiarities of the samples. The main sequence has maxima at $110-180$, $270-330$ and $400-450$ $h^{-1}$ Mpc, which limits their spacing to the range $110 < \lambda_p < 170$ $h^{-1}$ Mpc. The average value of $\lambda_p$ is $140$ $h^{-1}$ Mpc in a very good agreement with $130$ $h^{-1}$ Mpc found from the pencil-beam surveys.

Although we detect a periodic behavior of the correlation function, it is not clear whether there is a simple geometry in the distribution of matter. It has been shown that distributions based on Voronoi polygons give rise to a characteristic scale (Ikeuchi & Turner 1991), but they can not produce this scale and the observed small-scale amplitude of $\xi$ simultaneously (Williams, Peacock & Heavens 1991).

Our results set limits to the position of the maximum of the power spectrum $P(k)$. In order to produce the fluctuation amplitude and period found from the data, the spectrum must contain a distinct peak at $k_m = 0.048 \pm 0.005$ $h$ Mpc$^{-1}$ ($\lambda_m = 120 - 145$ $h^{-1}$ Mpc). This result is in agreement with upper limits for $k_m$ ($\lambda_m = 100 - 160$ $h^{-1}$ Mpc) found by different surveys, strongly confining the overall shape of the spectrum.

Financial support by the Max–Planck–Institut für Extraterrestrische Physik where a large part of this work has been performed is gratefully acknowledged. We thank Professor Joachim Trümper and Dr Rudolf Treumann for interesting discussions and Jaan Pelt and Enn Saar for their contribution.

This work was partially supported by Estonian Science Foundation grant 182 and International Science Foundation grant LDC 000.

---